\documentstyle[preprint,aps]{revtex}

\begin{document}
\title{From Kaluza-Klein theory to Campbell-Magaard theorem and beyond}
\author{C. Romero$^{a}$ and F. Dahia$^{b}$}
\address{$^{a}$Departamento de F\'{i}sica, Universidade Federal da Para\'{i}ba, Caixa%
\\
Postal 5008, 58051-970, Jo\~{a}o Pessoa, PB, Brazil \\
$^{b}$Departamento de F\'{i}sica, Universidade Federal de Campina Grande, \\
PB, Brazil\\
e-mail: cromero@fisica.ufpb.br}
\maketitle

\begin{abstract}
We give a brief review of recent developments in five-dimensional theories
of spacetime and highlight their geometrical structure mainly in connection
with the Campbell-Magaard theorem.
\end{abstract}

\ \ \ \ \ \ \ \ \ \ \ \ \ \ \ \ \ \ \ \ \ \ {\it ``Ainsi, dans la cosmologie
la g\'{e}ometrie s'organise}

\ \ \ \ \ \ \ \ \ \ \ \ \ \ \ \ \ \ \ \ \ \ \ {\it dans un cadre
id\'{e}alis\'{e} o\`{u} elle appara\^{i}t consubstantielle}

\ \ \ \ \ \ \ \ \ \ \ \ \ \ \ \ \ \ \ \ \ \ \ {\it \`{a} un ordre
pr\'{e}fabriqu\'{e} du monde.''}

\hspace{4.5cm}{\it (M.Novello, Cosmos et Contexte,1987)}

\medskip \vspace{2cm}

\section{Kaluza-Klein Theory}

The idea that our spacetime might have five dimensions and that some
observable physical effects could be attributed to the existence of a fifth
dimension has perhaps no better illustration than that provided by the
Kaluza-Klein theory \cite{1,2}. Indeed, in this theory a scheme is devised
in which the extra dimensionality of space is combined with curvature in
such a way that electromagnetic phenomena may be looked upon as a pure
manifestation of geometry. Although no new physical effect was to be
predicted by the theory, the very possibility of unification (even at a
mathematical level) between two different subjects such as gravity and
electromagnetism has had an appeal which endures until today. Of course
along the history of physics there have been many attempts to formulate a
unified field theory which would not resort to extra dimensions. In fact,
the fifth dimension assumption was basically the cause of Einstein's
reluctance to accept the plausibility of Kaluza-Klein theory \cite{3}.
Alternative ways to geometrize the electromagnetic field in the usual
four-dimensional spacetime may be found, for instance, in the work of Weyl,
who proposed a generalization of Riemannian geometry \cite{4}. More
recently, Weyl's approach has been revived by Novello through the WIST (Weyl
Integrable Space-Time) program, where interesting applications to Cosmology
have been discussed \cite{5}. The development of particle physics, on the
other hand, led to a ressurgence of interest in higher-dimensional field
theories, mainly as a possibility of unifying the long-range and short-range
interactions. Indeed, inspired by the old five-dimensional Kaluza-Klein
theory there appeared, around the sixties and seventies of the last century,
higher-dimensional theories such as eleven-dimensional supergravity and
ten-dimensional superstrings, all aiming at a unifying scheme \cite{6,7}.

\section{Induced Matter theory}

The original version of Kaluza-Klein theory assumes as a postulate that the
fifth dimension is compact (condition of cilindricity). Recently, however, a
non-compactified approach to Kaluza-Klein gravity, known as Induced-Matter
theory (IMT) has been proposed by Wesson and colaborators \cite{8}. The
basic principle of the IMT approach is that all classical physical
quantities, such as matter density and pressure, should be given a
geometrical interpretation \cite{9}. Moreover, it is asserted that only one
extra dimension should be sufficient to explain all the phenomenological
properties of matter. More specifically, IMT proposes that the classical
energy-momentum tensor, which enters the right-hand side of the Einstein
equations could be, in principle, generated by a pure geometrical means. The
theory also assumes that the fundamental five-dimensional space $M^{5}$, in
which our usual spacetime is embedded, should be a solution of the
five-dimensional vacuum Einstein equations

\[
R_{\alpha \beta }=0. 
\]

We shall not go into the details of the mathematical mechanism of
Induced-Matter theory (the interested reader is referred to \cite{8,9} and
the references therein). For our purposes here, suffice it to say that IMT
has a mathematical structure which can be better understood if it is
regarded as a spacetime embedding theory \cite{10}. In this context our
spacetime would correspond to a four-dimensional hypersurface locally and
isometrically embedded in a five-dimensional Ricci-flat manifold.

\section{Campbell-Magaard theorem}

Embedding theories are naturally subject to embedding theorems of
differential geometry. The claim that any energy-momentum can be generated
by an embedding mechanism may be translated in geometrical language as
saying that any semi-Riemannian four-dimensional manifold is embeddable into
a five-dimensional Ricci-flat manifold. At the time IMT was proposed it was
not at all apparent whether such proposition was true. Fortunately (at least
for the supporters of the theory), this is essentially the content of a
little known but powerful theorem due to Campbell \cite{11} and Magaard \cite
{12}, which asserts that any semi-Riemannian n-dimensional analytic manifold
can be locally and isometrically embedded in a semi-Riemannian
(n+1)-dimensional analytic manifold, where the Ricci tensor of the latter
vanishes \cite{13}. Campbell-Magaard's result has then acquired fundamental
relevance for granting the mathematical consistency of five-dimensional
non-compactified Kaluza-Klein gravity.

Local isometric embeddings of Riemannian manifolds have long been studied in
differential geometry. Of particular interest is a well known theorem
(Janet-Cartan) \cite{14,15} which states that if the embedding space is
flat, then the minimum number of extra dimensions needed to analytically
embed a Riemannian manifold is $d$ , with $0\leq d\leq n(n-1)/2$. The
novelty brought by Campbell-Magaard theorem is that the number of extra
dimensions falls drastically to $d=1$ when the embedding manifold is allowed
to be Ricci-flat (instead of Riemann-flat).

\section{ New embedding theories and the need for more general theorems}

The increasing attention given to the Randall-Sundrum models \cite{16,17} in
which the embedding manifold, i.e., the bulk, corresponds to an Einstein
space, rather than to a Ricci-flat one, has raised the question whether
Campbell-Magaard could be generalized and what sort of generalization could
be done. It was conjectured that if the Ricci-flatness condition were
replaced by the requirement of the embedding space being an Einstein space,
then a result similar to Campbell-Magaard theorem would hold. That this
conjecture is in fact a theorem was recently shown by Dahia and Romero \cite
{18}. Specific classes of embeddings, such as those of Einstein spaces into
Einstein spaces were established \cite{19}. This was the first extension of
Campbell-Magaard theorem and more were to come.

A following question to address was whether further extensions of
Campbell-Magaard theorem were possible. This led to an investigation of
embeddings in spaces that are sourced by dynamical matter fields. One of the
simplest forms of matter is that of a scalar field, so it was natural to
consider this case first. It was then proved that any n-dimensional analytic
Lorentzian or Riemannian space can be locally and isometrically embedded in
a (n+1)-dimensional analytic manifold generated by any arbitrary
self-interacting scalar field. As a corollary one can prove that any
Lorentzian or Riemannian n-dimensional analytic manifold can be embedded in
a (n+1)-dimensional space which is a solution of the vacuum Brans-Dicke
field equations \cite{20}.

In seeking higher levels of generalization one is led to consider the more
general situation of embedding spaces whose Ricci tensor is arbitrary.
Pursuing this idea a bit further a third extension of the Campbell-Magaard
was finally established. A theorem was proved which asserts that any
n-dimensional semi-Riemannian analytic manifold can be locally embedded in a
(n+1)-dimensional analytic manifold with a non-degenerate Ricci tensor which
is equal, up to a local analytic diffeomorphism, to the Ricci tensor of an
arbitrary specified space. As an application of this theorem, embeddings of
Ricci-flat spacetimes into five-dimensional Friedmann-Robertson-Walker
models were obtained in ref. \cite{21}.

To conclude we would like to point out that insofar as five-dimensional
embedding theories are metric it appears to be of relevance to allow the
embedding spaces to have different geometrical properties, which must
ultimately be determined by the dynamics of the theory in question.

Therefore, generalizations of the known embedding theorems might reveal
crucial in building new higher dimensional models. On the other hand, from
the standpoint of epistemology, it is rather illustrative to see how modern
theoretical physics can be a source of new ideas in mathematics and geometry.

\bigskip


\begin{references}
\bibitem{1}  Kaluza, T., Sitz. Preuss. Akad. Wiss. 33, 966 (1921)

\bibitem{2}  Klein, O. Z. Phys. 37, 895 (1926)

\bibitem{3}  Pais, A., The Science and the Life of Albert Einstein, Ch. 17
(Oxford University Press, 1982)

\bibitem{4}  Weyl, H., Space-time-matter (Dover, New York, 1952)

\bibitem{5}  Novello, M. Theoretical Cosmology in Proceedings of the VII
Brazilian School of Cosmology and Gravitation (Editions Fronti\`{e}res,1994,
Ed. M. Novello)

\bibitem{6}  Collins, P., Martin, A. and Squires, E., Particle Physics and
Cosmology (Wiley, New York, 1989)

\bibitem{7}  Appelquist, T., Chodos, A. and Freund, P., Modern Kaluza-Klein
Theories (Addison-Wesley, Menlo Park, 1987)

\bibitem{8}  Overduin, J.M. and Wesson, P.S., Phys. Rep. 283, 302 (1997)

\bibitem{9}  Wesson, P.S., Space-Time-Matter (World Scientific, 1999)

\bibitem{10}  Romero, C., Tavakol, R. and Zalaletdinov, R., Gen. Rel. Grav.
28, 365 (1995)

\bibitem{11}  Campbell, J. E., A Course of Differential Geometry (Oxford:
Claredon Press, 1926)

\bibitem{12}  Magaard, L. Zur einbettung riemannscher Raume in
Einstein-Raume und konformeuclidische Raume (PhD Thesis, Kiel, 1963)

\bibitem{13}  Dahia, F. Imers\~{a}o do espa\c{c}o-tempo em cinco
dimens\~{o}es e a generaliza\c{c}\~{a}o do teorema de Campbell-Magaard (PhD
Thesis, UFPb, Brazil , 2001)

\bibitem{14}  Einsenhart, L. , Riemannian Geometry ( Princeton University
Press, 1949), 143.

\bibitem{15}  Spivak, M. A Comprehensive Introduction to Differential
Geometry, vol. V (Publish or Perish, Houston, 1979)

\bibitem{16}  Randall, L. and Sundrum, R., Phys. Rev. Lett. 83, 3370 (1999)

\bibitem{17}  Randall, L. and Sundrum, R., Phys. Rev. Lett. 83, 4690 (1999)

\bibitem{18}  Dahia, F. and Romero, C., The embedding of the spacetime in
five dimensions: an extension of Campbell-Magaard Theorem, gr-qc/0109076

\bibitem{19}  Anderson, E. and Lidsey, J. E., Class. Quantum Grav. 18, 4831
(2001)

\bibitem{20}  Anderson, E. , Dahia, F. , Lidsey, J. E. and Romero, C.,
Embeddings in Spacetimes Sourced by Scalar Fields, gr-qc/0111094

\bibitem{21}  Dahia, F. and Romero, C. The embedding of the spacetime in
five-dimensional spaces with arbitrary non-degenerate Ricci tensor,
gr-qc/0111058
\end{references}
\end{document}